\documentclass[%
 reprint,
superscriptaddress,
 amsmath,amssymb,
 aps,
pra,
floatfix,
]{revtex4-2}

\usepackage{graphicx}
\usepackage{dcolumn}
\usepackage{bm}
\usepackage{multirow}
\usepackage{siunitx}
\usepackage{hyperref}
\usepackage[utf8]{inputenc}
\usepackage[T1]{fontenc}
\usepackage{mathptmx}

\usepackage{physics}

\newcommand{\ra}{\rangle}
\newcommand{\la}{\langle}

\newcommand*{\ipcfsu}{Institute of Physical Chemistry, Friedrich Schiller University Jena, Germany.}
\newcommand*{\abbe}{Abbe Center of Photonics, Friedrich Schiller University Jena, Germany.}
\newcommand*{\icmt}{Institute for Condensed Matter Theory and Solid State Optics, Friedrich Schiller University Jena, Germany.}
\newcommand*{\mpsp}{Max Planck School of Photonics, Jena, Germany.}
\newcommand*{\eqcontrib}{These two authors contributed equally.}

\begin{document}

\title{The impact of electron-electron correlation in ultrafast attosecond single ionization dynamics}

\author{Friedrich Georg Fröbel}
\thanks{\eqcontrib{}}
 \affiliation{\ipcfsu{}}
 \affiliation{\abbe{}}

\author{Karl Michael Ziems}
\thanks{\eqcontrib{}}
\affiliation{\ipcfsu{}}
\affiliation{\mpsp{}}

\author{Ulf Peschel}
\affiliation{\abbe{}}
\affiliation{\icmt{}}%

\author{Stefanie Gräfe}
\affiliation{\ipcfsu{}}
\affiliation{\abbe{}}
\affiliation{\mpsp{}}

\author{Alexander Schubert}%
 \email{E-mail: Alexander.Schubert@uni-jena.de}
 \affiliation{\ipcfsu{}}

\date{\today}

\begin{abstract}
The attosecond ultrafast ionization dynamics of correlated two- or
many-electron systems have, so far, been mainly addressed investigating
atomic systems. In the case of single ionization, it is well known that
electron-electron correlation modifies the ionization dynamics and
observables beyond the single active electron picture, resulting in effects
such as the Auger effect or shake-up/down and knock-up/down processes.
Here, we extend these works by investigating the attosecond ionization of 
a molecular system involving
correlated two-electron dynamics, as well as non-adiabatic
nuclear dynamics. Employing a charge-transfer molecular
model system with
two differently bound electrons, a strongly and a weakly bound electron,
we distinguish different pathways leading to ionization, be it direct
ionization or ionization involving elastic and inelastic electron
scattering processes. We find that different pathways result in a
difference in the electronic population of the parent molecular ion,
which, in turn, involves different subsequent (non-adiabatic) postionization
dynamics on different time scales.
\end{abstract} 

\maketitle

\section{\label{sec:intro}Introduction}
For many elementary processes in multi-electron systems, such as in photoionization, 
electron-impact ionization, the Auger effect, and other radiative processes, 
the impact of correlated electron-electron dynamics plays a crucial role
 \cite{Matveev82SovPhys}.
In the case of ionization, these electron correlations affect the state and 
the dynamics of the residual cation on an atto- and femtosecond timescale. 
As a consequence, the remaining bound electron(s) can be excited 
(shake-up/knock-up processes \cite{Sukiasyan12PRA}), relaxed (shake-down/knock-down), or even ejected subsequently (Auger process) \cite{Auger25JPR}.
Such time-resolved ionization dynamics of multi-electron systems have 
been investigated in various theoretical and experimental studies, for a 
review see for example~Ref.\,\citenum{Pazourek15RMP}. 

On the experimental side, the advent of ultrashort femto- 
or even attosecond pulses in the extreme ultraviolet (XUV)  regime being 
available either via table-top high-order harmonic generation or the newest 
generation of (X)FEL sources, for example, paved the way for 
the observation of these processes in real 
time, e.g.~employing the attosecond streaking technology 
\cite{Uiberacker07Nature, Corkum07NatPhys, Calegari16JPB,Zherebtsov11JPB,Ossiander17NatPhys}. 
However, resolving the details of these complex correlated many-body 
phenomena involving nuclear and electronic degrees of freedom 
still poses a challenge for computational simulations.

Most of the theoretical work has been focused on the helium atom 
as the simplest two-electron system
\cite{Hino93PRA,Ossiander17NatPhys,Pazourek12PRL,Sukiasyan12PRA}.
Effects which have been examined are the Wigner-Smith time delay 
and electron-electron correlation under the influence of an (infra\-red) laser field in the 
context of streaking spectroscopy \cite{Pazourek15RMP, Pazourek12PRL, 
	Sukiasyan12PRA, Kluender13PRA, Kazansky07JPB}. 
For molecules, the physics becomes even more complex, also due to the 
multi-scatterer nature compared to the centrosymmetric atoms.
Ning et al.~have investigated the simplest, prototypical molecule, H$_2^+$, 
and found pronounced interference effects (Cohen-Fano interferences, 
\cite{Cohen66PR}) caused by the two scattering centers \cite{Ning14PRA}. However, this prototype 
case inherently does not involve any multi-electron effects.
Other molecular systems investigated theoretically include endohedral complexes of type 
A@C$_{60}$ \cite{Deshmuk14PRA, Pazourek13Faraday} or, employing 
multi-configurational approaches, ionization and subsequent charge 
migration of small amino acids \cite{Remacle11PRA,  Hennig05JPCA, Kuleff07CP, Remacle06PNAS, Ayuso17PCCP, Lepine14NatPhot}. A few works have considered both, the correlated electron dynamics and 
the nuclear motion \cite{Despre15JPCL, Vacher17PRL, Vacher15PRA, Sun17JPCA}.

In order to reduce computational costs, numerical simulations 
commonly employ approximations such as the sudden approximation, 
frozen nuclear degrees of freedom, or the single-active electron 
approximation \cite{Kulander88PhysRevA}. 
Within the framework of the latter, one assumes that the dynamics of 
the active electrons is sufficiently fast so that the "inactive" electrons 
do no adapt within its time scale, i.e.~no electron correlated dynamics occurs. 
However, it has been shown by Awasthi et al.~that  electron correlation 
even upon XUV- or X-ray-induced ionization of multi-electron systems 
plays an important role and cannot be neglected \cite{Awasthi08PhysRevA,Awasthi10PhysRevA}.

\begin{figure}[t!]
\centering
    \includegraphics[width=8.5cm, trim=0 0 0 0, clip]{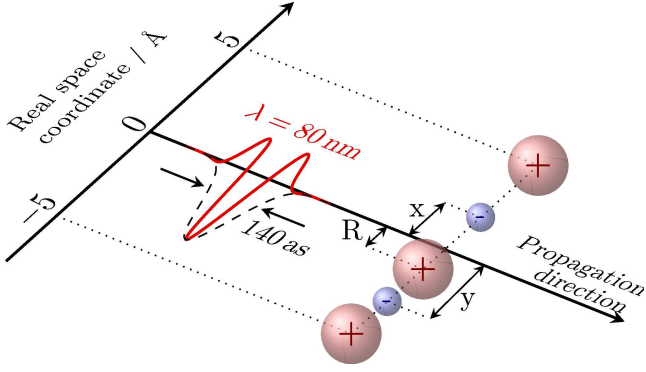}
    \caption{\label{fig:particleConf} Configuration of the extended 
    Shin-Metiu system: An ultrashort XUV pulse is used to ionize a linear 
    molecule aligned with the pulse's polarization axis. The molecule 
    consists of two fixed nuclei at $\pm5$\,\AA, two mobile electrons 
    with coordinates $x$ and $y$, and a movable central nucleus at $R$. 
    The mobile nucleus is initially localized at negative 
    $R$ values, whereas the two electrons reside on both sides. Thus, in 
    the electronic ground state, the electron at negative coordinates is 
    stronger bound than the electron located at positive values.}
\end{figure}

In this work, we investigate the effects of such electron-electron correlation
as well as non-adiabatic effects on the postionization dynamics of
a molecular charge transfer model system in a time-resolved picture.
Moreover, we aim to thoroughly distinguish the processes contributing 
to the electron-nuclear post-ionization dynamics.
The model system, which has been originally suggested by Shin and Metiu 
\cite{Shin95JCP, Shin96JPC}, has been extended by the group of Engel to 
include two electrons \cite{Erdmann04JCP2, Falge12JPCA}. This system, 
which will be introduced in more detail in Sec.\,\ref{sec:theory}, 
possesses, due to asymmetric initial conditions, one stronger and one 
weaker bound electron with anti-parallel spin. 
Interaction with an attosecond XUV pulse leads to 
the emission of one electron. 
We show that several processes occur on different timescales 
which have an impact on the electronic configuration of the residual molecular ion: 
(a) direct photoemission of either the 
weaker or the stronger bound electron, without passing the respective other 
electron, yet, affecting its quantum state due to an altered 
electrostatic environment (shake-up/shake-down processes); 
(b)  emission processes, where upon photoabsorption, the 
accelerated electron needs to ''pass`` the other electron. 
The latter process involves an immediate electron-electron interaction, which 
leads to both, inelastic and elastic scattering. 
As a result, the second electron is either excited into higher bound states 
(knock-up), relaxed into lower states (knock-down) \cite{Sukiasyan12PRA}, 
or adopts the momentum of the electron originally accelerated 
by the electric field and is emitted in its stead (''indirect`` photoemission); and
(c) non-adiabatic transition processes during the postionization dynamics.

The paper is organized as follows: After a short description of the 
model and the numerical methods utilized in Sec.\,\ref{sec:theory}, we 
will present the different pathways, analyzed additionally with the help 
of desymmetrized wave functions and restricted interactions of the electrons 
with the external electric field. The resulting dynamics is discussed by means 
of the final state-dependence in the residual ion in Sec.\,\ref{sec:results}. The 
paper ends with a summary and conclusion in Sec.\,\ref{sec:summary}.

\section{\label{sec:theory}Theory}
\subsection{\label{subsec:model}The full model system}

The model we apply in this work represents an extension to the one 
originally suggested by Shin and Metiu \cite{Shin95JCP, Shin96JPC}. In 
their work, a linear, one-dimensional charge-transfer model 
system was employed consisting of two fixed nuclei with charges 
$Z_1$ and $Z_2$ at the positions $\pm L$/2, one moving nucleus ($Z$) 
in between with coordinate 
$R$, and one electron, here with coordinate 
$y$, giving rise to the potential:
\begin{align}
V^\text{1e}(y,R&)={} \frac{e^2}{4\pi\epsilon_0}\left[\frac{Z_1 Z}{|L/2-R|}\!+\!
   \frac{Z_2 Z}{|L/2+R|}\!-\!\frac{Z\:\text{erf}(|R-y|/R_c)}{|R-y|} \right.\nonumber \\[6pt] 
   -&\left. \frac{Z_1\:\text{erf}(|L/2-y|/R_f)}{|L/2-y|} - 
    \frac{Z_2\:\text{erf}(|L/2+y|/R_f)}{|L/2+y|} 
\right],
\end{align} 
where the error functions (erf) describe a truncated Coulomb interaction 
between individual particles.
The truncation parameters $R_f$ and $R_c$ specify the interaction 
strength between the electron and the fixed nuclei and the mobile nucleus, respectively 
\cite{Shin95JCP, Shin96JPC}.

Here, we use an extension to this model introduced by Engel and coworkers, 
where a second electron, $x$, is added to the system \cite{Erdmann04JCP2, Falge12JPCA}.
The 
whole particle configuration is shown in Fig.\,\ref{fig:particleConf}. The system's potential takes on the form 
\begin{align}
   V^{\mathrm{2e}}(x,y,R) ={}& V^\mathrm{1e}(y,R) 
+ \frac{e^2}{4\pi\epsilon_0}\left[ - \frac{Z\:\text{erf}(|R-x|/R_c)}{|R-x|} \right.
   \nonumber \\[6pt]
    &- \frac{Z_1\:\text{erf}(|L/2-x|/R_f)}{|L/2-x|} - 
    \frac{Z_2\:\text{erf}(|L/2+x|/R_f)}{|L/2+x|} \nonumber \\
   &\left.+\frac{\text{erf}(|x-y|/R_e)}{|x-y|} \right], \label{eq:pot}
\end{align}
where $R_e$ scales the electron-electron interaction 
\cite{Erdmann04JCP, Falge12JPCL, Falge11JCP, Falge12JPCA, Falge17PCCP, Erdmann03JCP, Erdmann04JCP, Erdmann04EPJD}. 
The fixed nuclei have a distance of $L = 10$\,\AA{}.
Nuclear charges are $Z = Z_1 = Z_2 = 1$. 
All truncation parameters have been set to $R_f = R_c = R_e = 1.5$\,\AA{}, 
corresponding to the weak-coupling regime \cite{Falge17PCCP}.

\begin{figure}
  \centering
    \includegraphics[width=8.5cm, trim=0 0 0 0, clip]{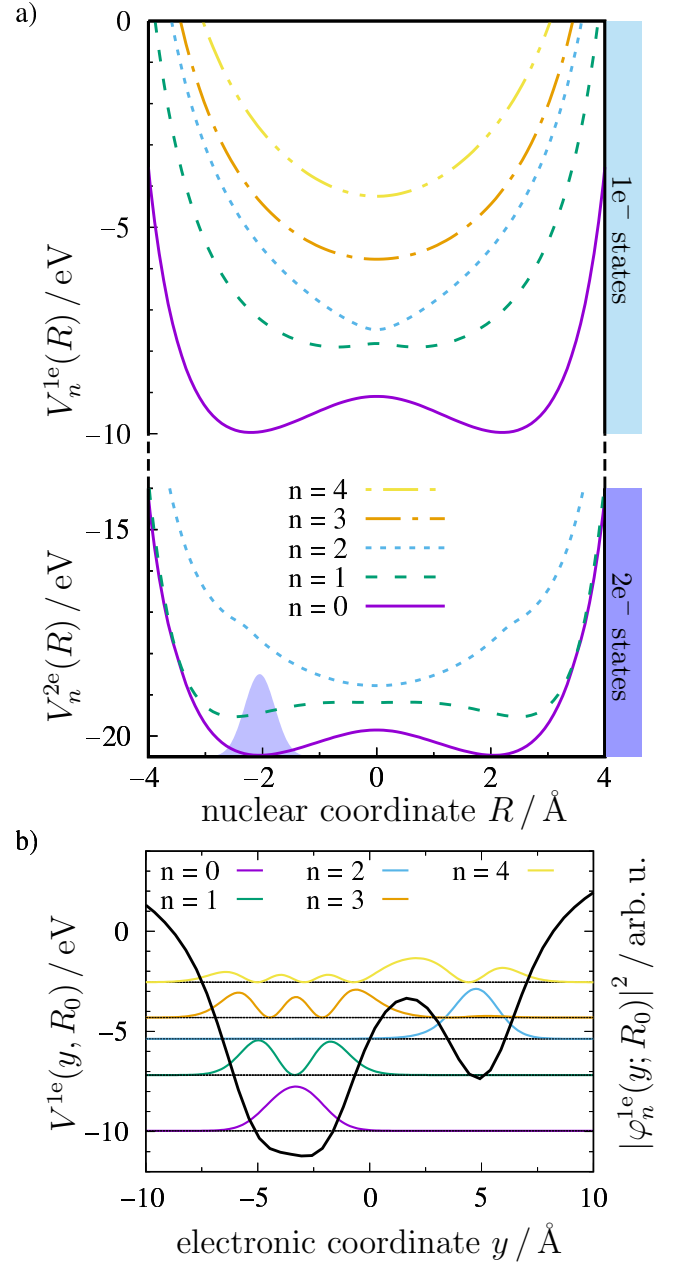}
    \caption{\label{fig:energycurves} (a) Potential energy curves 
    $V^{\text{1e,2e}}_n(R)$ for the lowest  electronic eigenstates 
    $\varphi^{\text{1e,2e}}_n$ of the one-electron (1e) and two-electron 
    (2e) systems. 
    The blue-shaded area depicts the initial nuclear wave packet $\chi(R)$
centered around the minimum at $R_0=-2.05\,\text{\AA{}}$.\newline
    (b) Potential energy $V^\mathrm{1e}(y,R_0)$ (solid black line) and first five electronic eigenfunctions of the single-electron 
    system, $\left|\varphi^{\text{1e}}_n(y;R_0)\right|^2$, at the initial, 
    near-equilibrium nuclear geometry $R_0$.
    }
\end{figure}

The three particle configuration of the model and its dynamics 
can be fully solved numerically. 
However, for interpretation, we calculate for the one-electron (1e) 
and the two-electron (2e) systems the (adiabatic) 
electronic eigenfunctions, $\varphi^{\text{1e}}_n(y;R)$ 
and $\varphi^{\text{2e}}_n(x,y;R)$, with eigenvalues $V^{\text{1e}}_n(R)$ and $V^{\text{2e}}_n(R)$, 
respectively, by solving the following eigenvalue equations:
\begin{align}
  \left[\frac{p_y^2}{2m_e}+ {V}^{\mathrm{1e}}(y,R) \right] \varphi^{\text{1e}}_n(y;R) 
  ={}& V^{\text{1e}}_n(R)\,\varphi^{\text{1e}}_n(y;R)\label{eq:eigenequation1}\\
  \left[\frac{p_x^2}{2m_e} + \frac{p_y^2}{2m_e} + {V}^{\mathrm{2e}}(x,y,R) \right] 
  &\varphi^{\text{2e}}_n(x,y;R) ={}\notag\\ &V^{\text{2e}}_n(R)\,
  \varphi^{\text{2e}}_n(x,y;R), 
  \label{eq:eigenequation2}
\end{align}
where $m_e$ is the electron mass and $p_x$ and $p_y$ refer to the electronic momenta.
For the two-electron case, the wavefunctions are symmetrized according to 
Pauli's principle and correspond to an anti-parallel spin configuration.
The obtained potential energy curves, $V_n^{1e/2e}(R)$, are shown in 
Fig.\,\ref{fig:energycurves}a for one (upper panel) and two bound electrons (bottom panel).
The one-electron model will be used to analyze the postionization dynamics of one
($y$) electron remaining in the parent ion after removal of the other ($x$) electron upon ionization. 
The first five electronic eigenfunctions of the single electron model, 
$\varphi_n^{\text{1e}}(y;R_0)$, are shown in Fig.\,\ref{fig:energycurves}b 
for an asymmetric nuclear configuration, $R_0=-2.05$\,\AA{}, corresponding
to the system initialization (see below).
Note, that among these states, for $n=0,1$, and 3, the electron is mostly 
localized on the left-hand side ($y<0$), i.e.~at the two close nuclei (in a 
strongly bound location), whereas for $n=2$ and 4 it is predominantly located
at the right hand side ($y>0$), i.e.~at the single nucleus (weakly bound).

The system interacts with an ultrashort attosecond XUV pulse which, 
using the dipole approximation and velocity gauge, results in the Hamiltonian:
\begin{align}
  {\mathcal{H}}(t) ={}&\frac{{P}^2}{2M} + \frac{{p}_x^2}{2m_e} + 
  \frac{{p}_y^2}{2m_e} + V^{\mathrm{2e}}(x,y,R) \notag\\ 
  {}&+e\,\mathcal{A}(t)\left(-\frac{Z}{M}{P}+\frac{p_x}{m_e} + 
  \frac{p_y}{m_e}\right),
  \label{eq:SE}
\end{align}
where $M$ is the proton mass and $P$ the nuclear 
momentum. 
The electric field of the ultrashort ionizing XUV pulse is described via 
its vector potential $\mathcal{A}(t)$ with a polarization along the
molecular axis:
\begin{equation}
\mathcal{A}(t) = -\frac{E_0}{\omega}\,g(t-T)\,\sin\big(\omega (t - T) + \phi\big).
\end{equation}
Here, $E_0 = -8.7$\,V/\AA{}~(or -0.169\,a.u.) is the electric field strength (corresponding 
to an intensity of 1.0$\times 10^{15}$ W/cm$^2$), $\omega$ the field's 
angular frequency, and $g(t-T)$ a Gaussian pulse envelope function 
centered around {$T = 0$}~fs with a full-width half-maximum (FWHM) of 
$140$~as (2.894\,a.u.). The angular frequency corresponds to a 
wavelength of $\lambda = 80$\,nm ($\hat{=} \; 15.5$\,eV or 0.570\,a.u.), which is 
sufficient to singly ionize the molecule through single photon absorption, 
see Fig.\,\ref{fig:energycurves}a. The spectral width of the attosecond 
pulse intensity is 18.4\,eV (0.676\,a.u.). 
The carrier-envelope phase (CEP) $\phi$ 
is set to zero, corresponding to a sine-shaped vector potential or an 
approximately cosine-shaped electric field. 

\subsection{\label{subsec:numericaldetails}Propagation and initialization}
The full system's wave function $\Psi(x,y,R,t)$ is represented on a 
three-dimensional grid with a range of \mbox{[-240,\,240]\,\AA{}} with 
1024 grid points in $x$- and $y$-direction, respectively, and of 
\mbox{[-4.99,\,4.99]\,\AA{}} with 128 points along the $R$-direction.
The time-dependent Schrödinger equation for the Hamiltonian defined in 
Eq.\,(\ref{eq:SE}) is solved numerically with a timestep of 5\,as using 
the split-operator technique \cite{FeitJCoPh82} and the FFTW\,3 library 
\cite{FFTW3} for Fourier transforms.  For the details on the numerics
please see our previous publications, e.g.~Ref.\,\citenum{Falge17PCCP}.
The simulation starts at $t_0=-4$\,fs, well before the XUV pulse enters 
the system. Reflection at the grid boundaries is suppressed
by multiplying $\Psi(x,y,R,t)$ at each timestep with
a splitting function \cite{Metiu1987} 
\begin{align}
f(x,y)={}&
\big[1+\mathrm{e}^{\zeta_1\,(|x|-\zeta_2)}\big]^{-1}\,
\big[1+\mathrm{e}^{\zeta_1\,(|y|-\zeta_2)}\big]^{-1}
\end{align}
with the parameters $\zeta_1=5.67$\,\AA{}$^{-1}$ and $\zeta_2=235$\,\AA{}.

\subsubsection{\label{subsubsec:fullsys}Full fermionic wave function}

The initial state is assumed to be a product state of the two-electron adiabatic 
electronic ground state, $\varphi_0^{\mathrm{2e}}(x,y;R)$, and a nuclear wave 
function 
\begin{equation}
\Psi(x,y,R,t_0)=\varphi_0^{\mathrm{2e}}(x,y;R)\,\chi(R). 
\label{eq:iniwf}
\end{equation}
The adiabatic electronic eigenstates are obtained by solving the field-free electronic 
Schrödinger equations, Eqs.\,(\ref{eq:eigenequation1}) and (\ref{eq:eigenequation2}), via the relaxation method \cite{Kosloff86CPL}.

The nuclear part of the initial wave function, $\chi(R)$, is assumed to be a 
Gaussian-shaped vibrational wave packet centered around the left local minimum 
of the double-well potential at $R_0\!=\!-2.05$\,\AA{} (see shaded area 
in Fig.\,\ref{fig:energycurves}a):  
\begin{equation}
\chi(R) = N_0\,e^{-\beta_R (R-R_0)^2}.
\label{eq:inichi}
\end{equation}
Here, $N_0$ serves as normalization constant and the 
width $\beta_R\!=\!7.14$\,\AA$^{-2}$. This way, the Gaussian closely resembles
the left-hand side of the vibrational ground state eigenfunction,
which is symmetric around $R=0$.
We note that the factorized, asymmetric initial state 
does not correspond to the total ground state of the system.
It rather corresponds to one of two energetically equal realizations.
This is a common situation, where
 the system resides in one potential well as encountered, for example, 
 in NH$_3$ inversion or isomerization processes.

\begin{figure}
\centering
    \includegraphics[width=8.5cm, trim=0 0 0 0, clip]{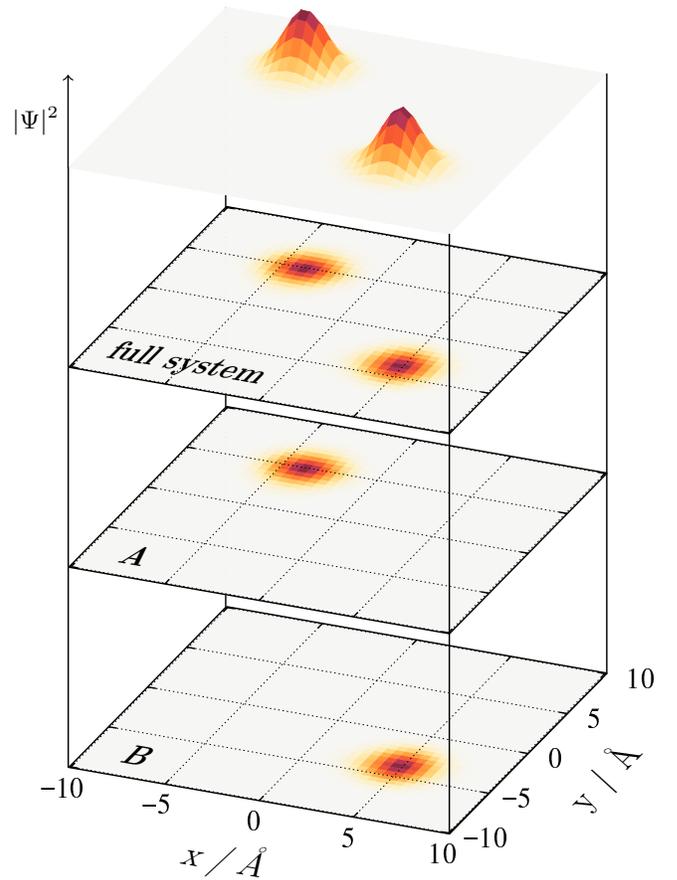}
    \caption{\label{fig:initalWF} Initial two-electron densities, 
    $\int |\Psi(x,y,R,t_0)|^2\,\mathrm{d}R$, entering the propagation. 
    Top plane: fully symmetric spatial wave function calculated via the 
    relaxation method. Descending from top to bottom: Projection onto the 
    2D plane of the full system and the artificial subsystems, (A) and (B), 
    respectively. }
\end{figure}

The initial two-electron densities, i.\,e.
$\int |\Psi(x,y,R,t_0)|^2\,\mathrm{d}R$,
is displayed in the top layer of Fig.\,\ref{fig:initalWF}.
Note that cuts through the spatial distribution of the electronic part -- given by the two-electron groundstate
wavefunction $\varphi^\text{2e}_0(x,y;R)$ -- approximately correspond 
to the one-electron functions $\varphi^{1e}_0(y;R)$ and $\varphi^{1e}_2(y;R)$.
The simulation results obtained for these initial conditions (and their relation to the
one-electron electronic eigenfunctions) are discussed 
in Sec.\,\ref{subsec:channels}. Please note that as we neglect any 
spin-dependent interaction, spin and spatial coordinates factorize. We 
therefore only consider the (symmetric) spatial part of the full wave 
function.

\subsubsection{\label{subsubsec:subsys}Artificial subsystems with 
distinguishable electrons}
For analysis purposes, the 
full wave function, Eq.\,(\ref{eq:iniwf}), is partitioned into two 
desymmetrized subsystems with 
\begin{subequations}
\begin{align}
\text{(A)~~}\psi_A(x,y,R,t_0) \equiv{}&\sqrt{2}\, \Psi(x,y,R,t_0)\,\Theta(y-x)
,\label{eq:psiA}\\
\text{(B)~~}\psi_B(x,y,R,t_0) \equiv{}& \sqrt{2}\, \Psi(x,y,R,t_0)\,\Theta(x-y)
,\label{eq:psiB}
\end{align}
\end{subequations}
respectively (Fig.\,\ref{fig:initalWF}, lower panels). In above equation,
 $\Theta(x)$ is the Heaviside step function.
These partial wave functions each describing one half of the full system, 
split along the $x\!=\!y$-diagonal. By doing so, 
the wave functions $\psi_A(x,y,R,t_0)$ and $\psi_B(x,y,R,t_0)$ vaguely resemble a wave function in 
Hartree-product form, because now $x$ and $y$ effectively 
describe identical, yet \textit{distinguishable} electrons. 
In the initial configuration of subsystem A (B) the $x$ ($y$) electron is strongly 
bound (with an approximate binding energy of 
$E_B^{\text{strong}}=V^\mathrm{1e}_2(R_0)-V^\mathrm{2e}_0(R_0)=15.1$\,eV), 
whereas the $y$ ($x$) electron 
($E_B^{\text{weak}}=V^\mathrm{1e}_0(R_0)-V^\mathrm{2e}_0(R_0)=10.5$\,eV)
is weakly bound. Technically, the abrupt cut-off of the wavefunction 
leads to a weak field-free ionization signal. 
This background signal is removed from the propagated wave function until 
$t=-250$\,as, i.e.~before the ionizing XUV pulse interacts with the system, by 
truncating $\Psi(x,y,R,t\leq-250\,\text{as})$ through multiplication with 
 $\Theta(25\,\text{\AA{}}-\abs{x})\,\Theta(25\,\text{\AA{}}-\abs{y})$.
The different ionization pathways revealed by these subsystems' dynamics 
are discussed in Sec.\,\ref{subsec:shakeup}.

\subsubsection{\label{subsubsec:redH}Restricted field interaction}
For analysis, further disentanglement of individual ionization pathways 
and their underlying intramolecular dynamics is achieved by artificially
 restricting the field interaction. To this end, simulations on subsystems A
  and B are performed with the modified Hamiltonian
\begin{align}\label{eq:modH}
 {\mathcal{H}}'(t) ={}&\frac{{P}^2}{2M} + \frac{{p}_x^2}{2m_e} + 
 \frac{{p}_y^2}{2m_e} + V^{\mathrm{2e}}(x,y,R) \notag\\ 
  {}&+e\,\mathcal{A}(t)\left(-\frac{Z}{M}{P}+\frac{p_{\xi}}{m_e}\right),
\end{align}
where $\xi\in\{x,y\}$ refers to either the $x$ \textit{or} the $y$ electron. 
Thus, field interaction is limited to a specific electron. This restriction 
allows us to distinguish direct and electron correlation driven photoionization
pathways.
Obtained results are presented and discussed in Sec.\,\ref{subsec:collision}.

\subsection{Identification of single ionization}
\label{subsec:ionanalysis}
To isolate fractions of the wave functions that belong to the singly 
ionized system, the photoelectron dynamics are devined via 
the ionization signal in regions far from the molecule using the mask function
\begin{equation}\label{eq:mask}
  c(x,y) = c_x(x) \cdot \big[1 - c_y(y)\big] 
\end{equation}
with
\begin{equation} \label{eq:cutoff}
 c_\xi(\xi) = 
  \begin{cases} 
   0 & \text{if } 0~<~\lvert \xi \rvert \leq \xi_c\\
   \sin^2\big(\frac{\lvert \xi \rvert-\xi_c}{\Delta\xi}\frac{\pi}{2}\big) 
   & \text{if } \xi_c < \lvert \xi \rvert \leq \xi_c\!+\!\Delta\xi \\
   1 & \text{if } \xi_c\!+\!\Delta\xi < \lvert \xi \rvert \leq 
 \xi_\mathrm{end}
  \end{cases},
\end{equation}
where $\xi\in\{x,y\}$ with corresponding $x_c = y_c = 25$\,\AA{}, 
$\Delta x =\Delta y = 10\,$\AA{}, and 
$x_\mathrm{end}=y_\mathrm{end}=240\,$\AA{} marking the 
end points of the simulation grid.
As a result, the mask $c(x,y)$ selects parts of the electronic wave 
functions 
\begin{equation}
\psi_{\text{out}}(x,y,R,t) \equiv c(x,y)\,\psi(x,y,R,t)
\end{equation}
at large $x$ and low $y$ coordinates, corresponding to the emission of the $x$ electron, 
whereas the other ($y$) electron 
remains bound at the parent molecular ion. 
Due to symmetry in the electronic coordinates, it is sufficient to only 
evaluate signals along the $x$ direction.
Note, that this mask is independent of $R$, as the nuclear part of the 
wave function remains well confined between the two outer nuclei.
Further segmentation of $\psi_\text{out}(x,y,R,t)$ into subregions $S$ will 
be introduced in Sec.\,\ref{subsec:channels}.

\section{\label{sec:results}Results and Discussion}
\subsection{\label{subsec:channels}Fully correlated fermionic wave function}
Interaction of the initial state with the ultrashort XUV pulse 
induces electron dynamics within 
the (non-ionized) two-electron system and leads to single ionization.
While for atomic systems, an ionization signal can be extracted 
via projection of the wave function onto a set of 
Coulomb waves \cite{Pazourek12PRL},
such an approach is not feasible for the multi-centered 
potential of the molecular model employed here.
Instead, we 
remove the two-electron components for the first 35 two-electron states, for which
both electrons are bound, 
from the total wavefunction. Then we project the remainder (containing only a single bound electron)
 onto the basis spanned by the electronic eigenfunctions 
$\{\varphi_n^{\text{1e}}(y;R)\}$ of the one-electron system \cite{Sukiasyan12PRA}:
\begin{align}
&P_n(t) ={} \iint \left|{\int \varphi^{\text{1e}}_{n}(y;R)\,
\Psi^{\text{1e}}(x,y,R,t)~\mathrm{d}y}\right|^2 \mathrm{d}x~\mathrm{d}R,
\end{align}
using the following defintions:
\begin{align}
&\Psi^{\text{1e}}(x,y,R,t) \equiv{} \Psi(x,y,R,t) - \sum_{m=0}^{34} a^\text{2e}_m(t)\,
\varphi^{\text{2e}}_m(x,y;R),\\             
&a^\text{2e}_m(t) \equiv{} \iiint{ \varphi^{\text{2e}}_{m}(x,y;R)\,\Psi(x,y,R,t)~
\mathrm{d}x}~\mathrm{d}y~\mathrm{d}R.
\end{align}

\begin{figure}[htbp]
  \centering
    \includegraphics[width=8.5cm, trim=0 0 0 0, clip]{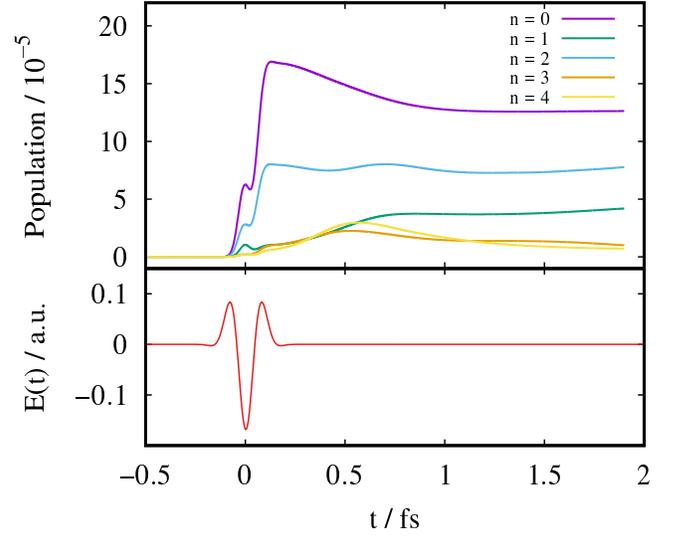}
    \caption{\label{fig:2eproj} 
   Population of the single-electron states upon projection onto the basis spanned by
 $\{\varphi_n^\mathrm{1e}(y;R)\}$ after removal of all two-electron 
states (upper panel) and temporal profile of the XUV pulse's electric field (lower panel).
    }
\end{figure}

The populations $P_n(t)$ are shown in Fig.\,\ref{fig:2eproj} upper panel. 
Note, that in the basis considered here, only contributions are obtained, where the 
$y$ electron remains bound, while the $x$ electron is ejected. 
Projection onto $\{\varphi^\text{1e}_n(x;R)\}$ yields the same results for reversed roles.
One finds that the first five one-electron states are considerably populated through 
the ultrashort XUV pulse (Fig.\,\ref{fig:2eproj} lower panel) around $t=0$.

For comparison, within the single active electron approximation, the \emph{sudden} removal 
of one of the electrons would yield time-independent one-electron state occupations 
obtained by projection of $\varphi_0^\text{2e}(x,y;R)$ onto $\{\varphi^\text{1e}_n(y;R)\}$:
\begin{align}
b^{\text{1e}}_n \equiv{}{}& \iint \big|\int\varphi^{\text{1e}}_{n}(y;R)\,\varphi^{\text{2e}}_0(x,y;R)~
\mathrm{d}y \big|^2\,\mathrm{d}x~\mathrm{d}R.
\end{align}
For the initial conditions considered here, only the states $n=0$
 (ejecting the weakly bound electron) and $n=2$ (ejecting the strongly bound
 electron) would be populated significantly, leaving the remaining electron more tightly bound in
the molecular system (\emph{shake-down} process \cite{Ossiander17NatPhys}). 
The occupation of higher one-electron states, 
i.e.~$n=$1,3, and 4, corresponding to a \emph{shake-up} process \cite{Ossiander17NatPhys},
would be approximately two orders of magnitudes lower ($b^{\text{1e}}_1/b^{\text{1e}}_0=0.013$, $b^{\text{1e}}_3/b^{\text{1e}}_0=0.008$, $b^{\text{1e}}_4/b^{\text{1e}}_2=0.006$).

In contrast, in our simulation with fully correlated electrons, depicted in Fig.\,\ref{fig:2eproj}, 
these three states show significant occupations.
It is also noteworthy, that their population 
continues to rise after the XUV pulse has passed the system, while in particular 
the population of the one-electron groundstate ($n=0$) declines.
Since non-adiabatic transitions occur on a much longer timescale (see below), 
we trace these phenomena back to the continued interaction between the 
bound ($y$) and the ejected ($x$) electron on early timescales,
where a transition to higher bound states is associated with a \emph{knock-up} process
and one to lower states corresponds to a \emph{knock-down} process \cite{Sukiasyan12PRA}.

\begin{figure}
  \centering
    \includegraphics[width=8.5cm, trim=0 0 0 0, clip]{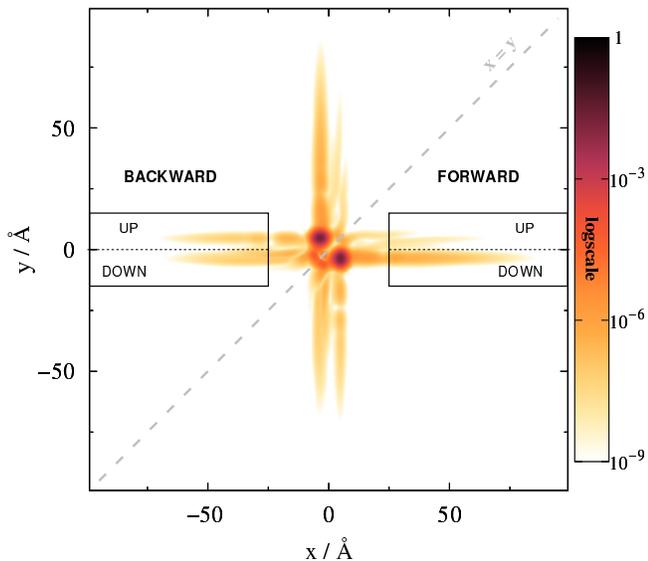}
    \caption{\label{fig:channel} Two-electron density, 
    $\int |\Psi(x,y,R,t=3\text{\,fs})|^2\,\mathrm{d}R$, of the fully 
    antisymmetric system (mirror symmetry w.r.t.~the $x=y$ diagonal) 
    3\,fs after ionization with an 140\,as XUV pulse centered at $T = 0$\,fs. 
    Four different ionization channels can be distinguished, here illustrated 
    for the emission of the $x$-electron ($x$ direction): 
    Emission occurs either in \emph{forward} 
    ($x>+25$\,\AA{}) or 
    \emph{backward} ($x<-25$\,\AA{}) direction for positive (\emph{up})
    or negative (\emph{down}) positions of the remaining $y$ electron. 
    The corresponding channels along the $y$-axis are equivalent.}
\end{figure}

In the following, we aim to disentangle and quantify the various correlation-induced 
processes that occur during different ionization pathways. 
To this end, we will evaluate the postionization dynamics.
Fig.\,\ref{fig:channel} depicts a snapshot of the two-electron density, 
$\int \abs{\Psi(x,y,R,t)}^2\,\mathrm{d}R$, of the full antisymmetric 
system at $t = 3$\,fs after XUV-pulse interaction. 
As can be gathered, the dominant part of the wave function remains around 
the origin (corresponding to the non-ionized part of the system), with minor 
parts being delocalized into the $x$ or $y$ direction. 
These four double-stripe structures, where either $x$ or $y$ coordinate 
stays localized within $\pm$10\,\AA{}, represent different single ionization processes. 
Electron densities in regions of high values of both coordinates, 
$|x|$ and $|y|$, simultaneously, corresponding to 
double ionization, are approximately four orders of magnitudes 
lower due to the much larger energy threshold for double ionization. 
Consequently, such contributions are not visible
in Fig.\,\ref{fig:channel}.
In the following, we will concentrate on the single ionization dynamics occurring during 
the interval of 1 to 20\,fs after pulse arrival.

Naturally, the electron densities are 
symmetric with respect to the $x\!=\!y$-diagonal.
It is therefore sufficient to restrict the analysis to 
electron densities emitted along one axis. 
Here we chose the $x$ axis and ascribe the labels \emph{forward}/\emph{backward} for positive/negative 
values in $x$.
An apparent feature of the single ionization channels is the occurrence 
of two ionization pathways in every direction. 
This indicates that the remaining electron eventually stays at different potential 
minima (around $R=0$), 
for which we introduce the labels \emph{up/down} for positive/negative values in $y$, 
respectively, as indicated in Fig.\,\ref{fig:channel}.
The electron densities in these four distinct ionization channels 
differ from each other in shape and amplitude.
To distinguish the underlying processes, an evaluation region 
is defined according to Eqs.\,({\ref{eq:mask}}) and (\ref{eq:cutoff}) 
and further separated into subregions $S$ 
according to the ascribed labels,
allowing us to collect and analyze the emitted wave function 
$\psi^S_\text{out}(x,y,R,t)$ of each channel separately. 
As the region of ionization is defined for \mbox{$\abs{x} \geq 25$\,\AA{},} 
no emission signal is detected until \mbox{$t \approx 1$\,fs}, when the 
fastest components of the ionized wave function enter the evaluation region.
Ionization signal, i.e.~$I_S(t)= \iiint \abs{\psi_\text{out}^S(x,y,R,t)}^2\,
\mathrm{d}x\,\mathrm{d}y\,\mathrm{d}R$, 
builds up within a subregion $S$ mainly over the period of $\sim$5\,fs.
The build up is traced back to a kinetic energy distribution whose 
central 80\%{} lie between 0.3 and 4.5\,eV with a maximum at 1.4\,eV.
After 5\,fs the overall probability to find both electrons in each subregion 
continues to increase only slightly 
due to parts of the wave packet with lower kinetic energy entering the subregion. 
The respective probabilities at 
5\,fs are $3.9\times10^{-6}$, $1.1\times10^{-4}$, $1.8\times10^{-5}$, 
and $2.8\times10^{-5}$ for the \emph{forward-up}, \emph{forward-down}, 
\emph{backward-up}, and \emph{backward-down} subregion, respectively.

\begin{figure}
\centering
    \includegraphics[width=8.5cm, trim=0 0 0 0, clip]{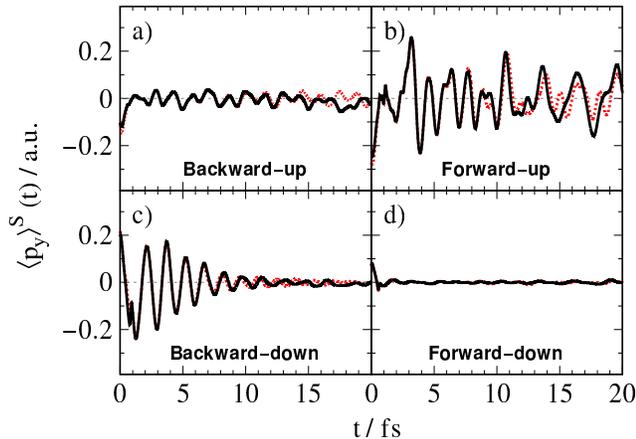}
    \caption{\label{fig:py} Time-resolved momentum expectation values 
    $\la p_y\ra^S (t)$  of the remaining $y$ electron (black solid lines) after ionization of 
    the full system for the different ionization channels, see Fig.\,\ref{fig:channel}, as indicated.
    For comparison, results from a simulation with a frozen nuclear configuration are added (red dotted lines).
    The expectation value $\la p_y\ra^S (t)$ 
    of the remaining bound electron $y$ serves as an indicator for electron scattering 
    during the emission process.
    }
\end{figure}

Fig.\,\ref{fig:py} shows the time evolution of the average momentum, 
$\la p_y\ra^S(t)$, of electron $y$, which remains bound in the molecular 
ion after electron $x$ has been emitted.
The quantity $\la p_y\ra^S(t)$ is calculated via the density distribution 
for the remaining $y$ electron, $\rho_S(p_y,t)$  by integrating over
$\psi^S_{\text{out}}$ in each individual channels $S$: 
\begin{equation}
   \la p_y\ra^{S}(t) = \frac{ \int p_y\,\rho_S(p_y,t)~\dd p_y}{\int \rho_S(p_y,t)~\dd p_y},
\end{equation}
where
\begin{equation}
    \rho_S(p_y,t) = \iint \left|
    \widetilde{\psi}_{\text{out}}^S(p_x,p_y,P,t)\right|^2~\dd p_x~\dd P, \\
\end{equation}
Here, $\widetilde{\psi}_{\text{out}}^S$ is the Fourier-transformed wave function $\psi^S_{\text{out}}$.
The average momentum $\la p_y\ra^{S}(t)$ illustrates that all four regions 
differ in the dynamics induced in the parent ion.

In the following, we will interpret the different dynamics seen 
in Fig.\,\ref{fig:py}a--d based on the leading contributions 
to photoemission into each evaluation region. However, 
there are further contributions to each channel, which 
will be isolated and investigated in Secs.\,\ref{subsec:shakeup} 
and \ref{subsec:collision}.

We interpret the observations as follows:
In the \emph{forward-down} channel, Fig.\,\ref{fig:py}d, the signal stems 
primarily from the \emph{direct emission} of the weakly bound electron on the 
right-hand side of the molecule 
towards positive $x$ values, i.e.~without passing the parent ion (and in particular the other electron) first. 
The strongly bound electron therefore remains on the left-hand side and is 
hardly affected by the ionization dynamics 
which is reflected in the nearly constant momentum expectation value of the remaining electron.
In contrast, the oscillating signal in the \emph{forward-up} channel, Fig.\,\ref{fig:py}b, 
can be primarily traced back to the strongly bound 
electron at negative $x$ values being released towards positive $x$ 
values, such that it first passes the parent ion and the other electron. 
Its inelastic scattering with the remaining, weakly bound electron 
induces oscillations of the latter being reflected by the strong 
time dependence of the remaining electron's average momentum.
We note that the temporal behavior significantly changes for times $t>10$\,fs, 
if the nuclear configuration is frozen during the simulation (red dotted lines)
suppressing non-adiabatic transitions.

A similar situation but with reversed roles can be seen for emission into 
the \emph{backward} direction: Here, the \emph{down} channel, 
Fig.\,\ref{fig:py}c, 
corresponds to emission of the weakly bound electron after passing the 
strongly bound one. 
Thereby, inelastic scattering leads to a regular oscillation in the residual electron's 
average momentum $\la p_y\ra(t)$. 
Fourier analysis of this oscillation occurring between $t=0$ and 10\,fs 
indicates that the corresponding energy of 2.89\,eV can be assigned to the energy gap 
between the electronic ground and first excited state of the one-electron system (2.78\,eV at $R_0$), 
see Fig.\,\ref{fig:energycurves}a. 
Thus, upon ionization, an electronic wave packet in the residual molecular ion is 
excited oscillating around the left well's minimum.

The \emph{backward-}\emph{up} channel, Fig.\,\ref{fig:py}a, on the other hand, corresponds 
to the direct emission of the strongly bound electron without passing the parent ion.
A weak response towards a negative average momentum of the remaining electron, 
can be seen, in particular between 17 and 20\,fs, 
despite the absence of an immediate interaction between the escaping and the 
remaining electron. 
This feature is not present if the simulation is performed with a frozen nuclear configuration 
(dotted red lines). It originates in the induced nuclear dynamics leading 
to non-adiabatic transitions (intramolecular charge transfer), which will be discussed 
in more detail in Sec.\,\ref{subsec:collision}.

\subsection{\label{subsec:shakeup} Reduced wave function: 
Distinguishable electrons}

\begin{figure*}[htbp]
\centering
    \includegraphics[width=15.9cm]{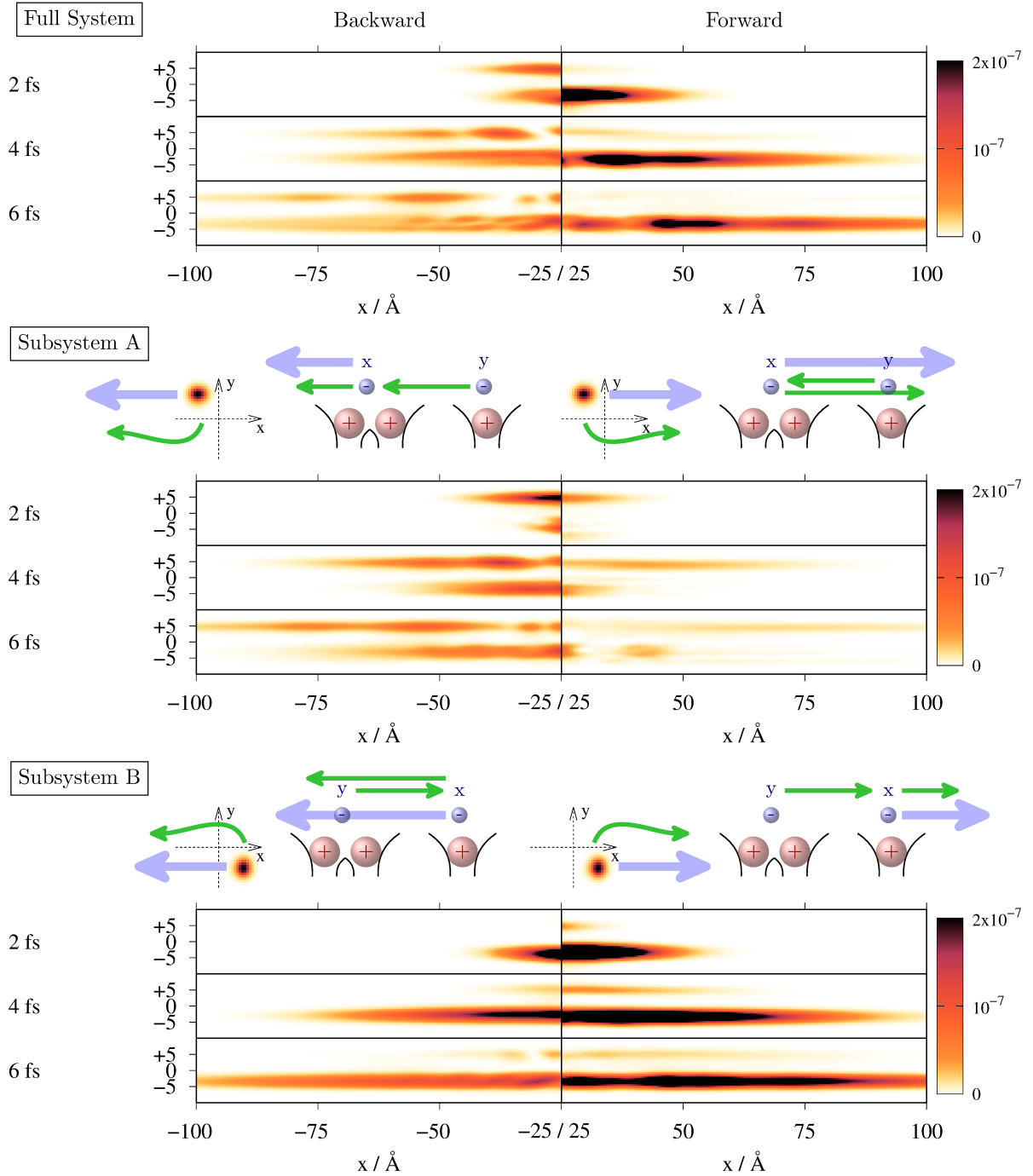}
    \caption{\label{fig:contributions} Snapshots of the integrated 2D 
    electronic density $\int |\Psi(x,y,R,t)|^2\,\mathrm{d}R$ at $t=$ 
    2, 4, and 6\,fs after ionization of the full system (top panel) and the two 
    subsystems, $\psi_A$ (middle panel) and $\psi_B$ (bottom panel), with 
    headers illustrating the ionization process (2D view of initial density 
    distribution in the configuration space and 1D view in the real 
    coordinate space). 
    Top: Ionization dynamics of the full system.
    Middle: Subsystem A with its initial wave packet centered at 
    negative $x$ and positive $y$ values. 
    The \emph{up} pathways 
    primarily contribute to the total ionization signals (blue arrows) 
    as opposed to the \emph{down} pathways (green arrows). The initially 
    strongly bound $x$-electron (negative $x$-values) is emitted and the 
    initially weakly bound $y$-electron (positive $y$-values) stays. 
    Bottom: Subsystem B with its initial wave packet centered at positive 
    $x$ and negative $y$ values: Here, the \emph{down} pathways primarily 
    contribute to the total ionization signals. 
    The initially weakly bound $x$-electron (positive $x$-values) is 
    emitted while the initially strongly bound $y$-electron 
    (negative $y$-values) stays. Oversaturation in the colorbars' range 
    is used to highlight substructures in the probability density as well 
    as small contributions in the non-dominating pathways. The largest 
    occurring values of the electronic density in \emph{forward} direction are: 
    (A) $8.3 \times 10^{-8}$, (B) $1.3 \times 10^{-6}$; and in \emph{backward} 
    direction: (A) $2.4 \times 10^{-7}$, (B) $4.0 \times 10^{-7}$.
    }
\end{figure*}

The previous analysis provided a first intuitive picture of intramolecular 
scattering effects in the course of ionization. However, as 
electrons are indistinguishable, the roles of emitted and remaining electrons 
during the scattering process cannot clearly be identified.
To this end, the artificially truncated wave functions $\psi_A$ and $\psi_B$, 
see Eqs.\,(\ref{eq:psiA}) and (\ref{eq:psiB}), are employed as initial 
conditions, 
thus rendering the two electrons distinguishable, see 
Sec.\,\ref{subsubsec:subsys}. 
This way, electron-electron correlation originating from the antisymmetry 
of the wave function and interference effects between the two distinct 
initial density distributions (localized near 
$x=\pm5$\,\AA{} \& $y=\mp5$\,\AA{}) are neglected.
However, a comparison of the probabilities to find the particles in the evaluation 
regions, $I_S(t)$, between the full system and the sum of the subsystems A and B 
shows very good agreement, indicating that for the present system these 
types of correlation effects are of minor importance. 

In Fig.\,\ref{fig:contributions} the time-dependent two-electron densities 
of the full system with two indistinguishable electrons (upper part) and the 
subsystems, A (middle part) and B (lower part) with distinguishable electrons,
are shown for an area corresponding to the emission of 
electron $x$ in \emph{backward} (left panels) and \emph{forward} direction
 (right panels), while electron $y$ remains bound to the parent ion. 
Above the two lower panels, a schematic picture indicates different 
underlying processes (blue/green arrows) in the $x,y$-configuration 
space (left) and the one-dimensional coordinate space (right). 
The thick blue arrows correspond to the four main contributions, 
i.e.~photoemission with and without intramolecular electron-electron 
scattering, discussed 
in Sec.\,\ref{subsec:channels}, where either the strongly (A) or the 
weakly bound electron (B) interacts with the electromagnetic field and
 is released to either side of the molecule.

The first electron wave packet components enter the evaluation region at 
$x=\pm25\,$\AA{} between 1 and 2\,fs after the interaction with the ionizing pulse. 
At this instant, electron density is mostly found in the \emph{backward-up}
channel for subsystem A and in the \emph{forward-down} channel for 
subsystem B (blue arrows) corresponding to direct photoemission from
the side of the molecule closest to the respective subregion.
Additionally, a strong slightly delayed signal can be noted 
 stemming from an emission into the opposite direction 
(blue arrows, A: \emph{forward-up}, B: \emph{backward-down}), 
corresponding to photoemission channels involving intramolecular 
electron-electron scattering, i.e.~the $x$ electron first passing the $y$ electron
before beind finally emitted. 

The A/B distinction reveals additional signals
with a smaller but still significant probability appearing in the \emph{down} 
(A) channels and -- to an even smaller extent -- in the \emph{up} (B) 
channels, which are not visible in Fig.\,\ref{fig:channel} due to the larger 
amplitude of the more dominant signals (blue arrows). 
Their appearance reveals a correlated motion between the two electrons  
(indicated by green arrows), where the remaining $y$ electron is relocated 
(intramolecular charge-transfer) either prior to the
photoelectron emission of the $x$ electron or afterwards. 
This question is addressed in the following section, \ref{subsec:collision}, 
by further dissecting the ionization pathways through restricting the 
electrons' interaction with the electric field.

\subsection{\label{subsec:collision}Restricted electron-field interactions}
To further investigate the intramolecular dynamics during and after the 
electron emission process, we perform simulations of the subsystems A and 
B, i.e.~using distinguishable electrons, and restrict the interaction of 
the electromagnetic field to either the $x$ (ejected) or  $y$ (remaining) electron by employing 
the modified Hamiltonian ${\mathcal{H}}'(t)$ defined in Eq.\,(\ref{eq:modH}).
This way, the absorption process of a system with distinguishable electrons
 is strictly limited to one specific single electron.

\begin{figure*}[tbp]
\centering
    \includegraphics[width=17cm]{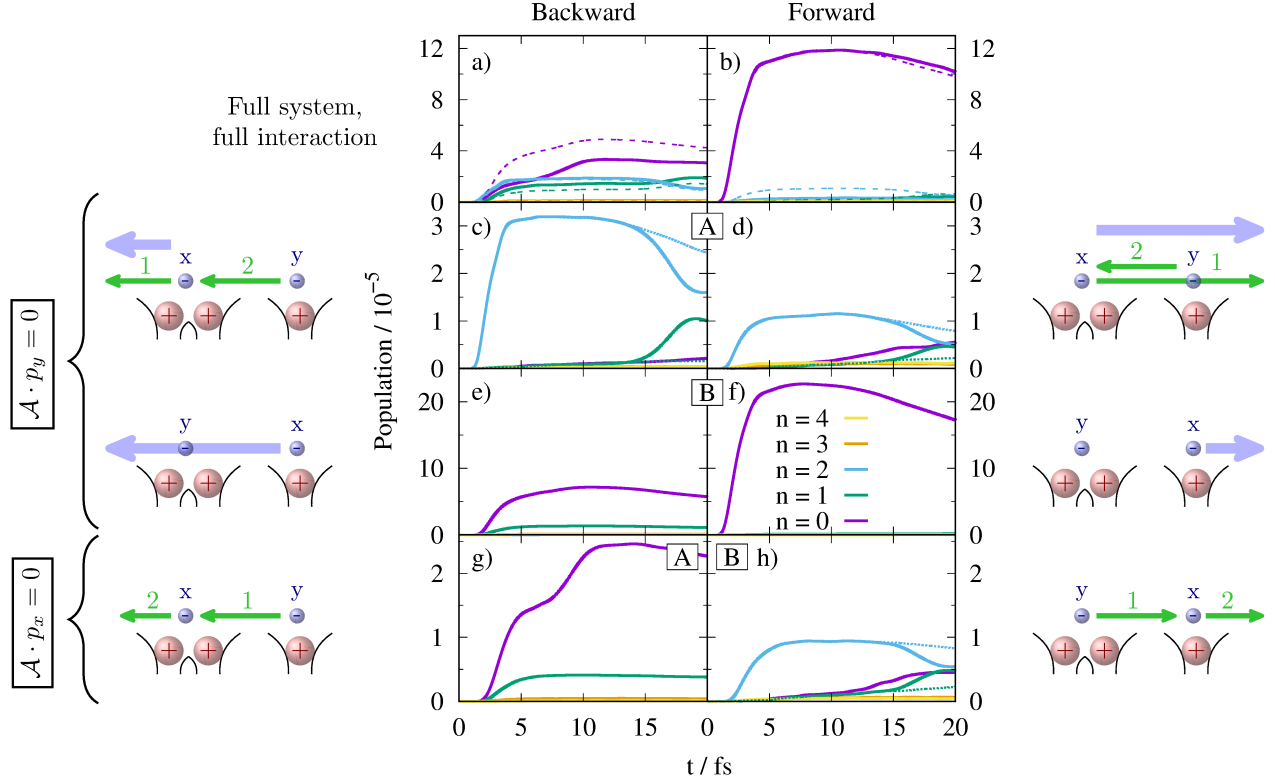}
    \caption{\label{fig:pop} Evolution of the population of the 
    electronic states by the remaining bound $y$-electron calculated 
    via Eq.\,(\ref{eq:pop}), after the XUV pulse ionizes the two-electron
    system at $t=0$ fs.
    The upper panels, (a) and (b), contain the dynamics of the full 
    antisymmetric system, whereas lower panels, (c)--(h), show the dynamics 
    of distinguishable electrons selectively interacting with the electric 
    field as indicated. 
    Dashed lines in panels (a) and (b) correspond to the sum of all
     individual pathways with restricted field interaction displayed 
     throughout panels (c)--(h).
    The dotted lines in panels (c), (d), and (h) represent evolution for a molecule 
    with fixed core position.     
    A comprehensive list of all processes can be found in Tab.\,\ref{tab:dichotomy}.
    }
\end{figure*}

The ionization wave function, $\psi^S_\text{out}(x,y,R,t)$, i.e.~the part 
of the wave function entering the analysis region defined via the mask 
function, Eq.\,(\ref{eq:cutoff}), is projected onto the set of adiabatic eigenfunctions
 $\{\varphi^{\text{1e}}_{n}(y;R)\}, n\in\{0,\dots,4\}$ of the one-electron system, 
 obtained from Eq.\,(\ref{eq:eigenequation1}) and shown in 
 Fig.\,\ref{fig:energycurves}b exemplarily for $R=R_0$. 
Note that these states differ in the electron's spatial distributions: 
while for states with the quantum numbers $n\in\{0,1,3\}$, 
the $y$ electron is located on the (strongly bound) left-hand side, 
for $n\in\{2,4\}$ electron distribution is predominantly 
found on the (weakly bound) right-hand side. 
Thus, a transition 
between states from different sides
 corresponds to an intramolecular charge transfer.
The population $P^{S'}_n(t)$ of the $n$th one-electron state by the $y$ electron 
is calculated as
\begin{align}
P^{S'}_n(t) = \iint \left|{\int \varphi^{\text{1e}}_{n}(y;R)\,
\psi^{S'}_{\text{out}}(x,y,R,t)~\mathrm{d}y}\right|^2 \mathrm{d}x~\mathrm{d}R, \label{eq:pop}
\end{align}
where the domain $S'$ limits the $x$ integration to either positive or 
negative values corresponding to the \emph{forward} ($S'=\mathrm{fwd}$) or \emph{backward} ($S'=\mathrm{bwd}$)
channel, respectively, but does not distinguish between \emph{up-} and 
\textit{down-}channels anymore.
Thus, $P^{S'}_n(t)$ are the populations of the single-electron states 
in the evaluation region, i.e.~in the molecular parent ion after photoionization. 
Sketches next to the panels illustrate the dominant ionization 
pathways with labels '1' and '2' indicating the temporal order.

The populations are shown in Fig.\,\ref{fig:pop} in \emph{backward} 
(left panels, ${x<0}$) and \emph{forward} direction (right panels, ${x>0}$) 
for the fully antisymmetric wave function (top panels, a and b) and the 
subsystems A and B as indicated (lower panels, c--h).
The rise in the period of 1 to 5\,fs in all panels corresponds 
to the main part of the wave packet 
entering the evaluation region. 
In the subsequent time evolution, the overall probability within the evaluation 
region remains mostly constant leading to distinct plateau regions in $P^{S'}_n(t)$.
However, small contributions with low kinetic energy (of the emitted $x$ electron) 
continue to enter the evaluation region at later times, while components with high 
kinetic energy are removed at the grid boundaries. 

\begin{table*}[hbtp]
\caption{Various in parts interdependent processes occur during
photoionization and contribute to the signal in different evaluation regions
for the two subsystems A (with a strongly bound $x$ electron and weakly
bound $y$ electron) and B (vice versa). The build up time describes the time frame 
of the initial wave packet entering the analyzing region (defined by signal above the 
noise level) until a stable population is reached. $P^{S'}_n(t)$ in the last column 
represents the maximum population of this process.
}
\label{tab:dichotomy}
\begin{tabular}{l|l|c|c|cl|l|c|S[table-format=2.2]}
\hline\hline
& Channel $S$ &$x$ & $y$ &\multicolumn{2}{l|}{Process} & 
Fig.\,\ref{fig:pop} & \multicolumn{1}{c|}{build up time / fs } & 
\multicolumn{1}{c}{$P^{S'}_\mathrm{n}$/$10^{-5}$}\\
\hline
\multirow{8}{*}{A} &  \emph{backward-up}& $<-25$\,\AA{} & $>0$ &\textbf{III}&\textbf{direct emission} of the strongly bound electron & c, blue  & 0.7 - 5.5 & 3.1\\ \cline{2-9}
& \multirow{3}{*}{\emph{backward-down}}& \multirow{3}{*}{$<-25$\,\AA{}} & \multirow{3}{*}{$<0$} &\multirow{2}{*}{\textbf{VI}}&\textbf{indirect emission} via elastic collision with charge transfer  & \multirow{2}{*}{g, green} & 1.2 - 14.0 & 2.5 \\
&&&&& and subsequent \textbf{knock up} (inelastic scattering) & & 1.7 - 8.2 & 0.4 \\ \cline{5-9}
&&&&\textbf{IX}& charge transfer via \textbf{non-adiabatic transition} (following \textbf{III}) & c, green &11.2 - 19.0 & 1.0 \\  \cline{2-9}
&  \multirow{2}{*}{\emph{forward-up}}& \multirow{2}{*}{$>+25$\,\AA{}} & \multirow{2}{*}{$>0$} &\multirow{2}{*}{\textbf{IV}} & \textbf{scattered emission} of the strongly bound electron & \multirow{2}{*}{d, blue} &   1.0 - 5.9& 1.1 \\
&&&&& and subsequent \textbf{knock up} (inelastic scattering) & & 1.3 - 4.5 & 0.1 \\ \cline{2-9}
& \multirow{2}{*}{\emph{forward-down}}& \multirow{2}{*}{$>+25$\,\AA{}} & \multirow{2}{*}{$<0$}&\textbf{V}&\textbf{knock down} induced charge transfer (following \textbf{IV})  & d, green &6.2 - 16.2 & 0.5 \\  \cline{5-9}
&&&&\textbf{X}& charge transfer via \textbf{non-adiabatic transition} (following \textbf{IV}) & d, green & 11.2 - 19.0 & 0.5 \\
 \hline
\multirow{8}{*}{B} &  \emph{backward-up}& $<-25$\,\AA{} & $>0$ & \multicolumn{2}{l|}{Process not visible} & --- & ---& 0.0\\ \cline{2-9}
& \multirow{2}{*}{\emph{backward-down}}& \multirow{2}{*}{$<-25$\,\AA{}} & \multirow{2}{*}{$<0$} & \multirow{2}{*}{\textbf{II}} & \textbf{scattered emission} of the weakly bound electron & \multirow{2}{*}{e, blue} & 0.8 - 7.2 & 6.6 \\
&&&&& and subsequent \textbf{knock up} (inelastic scattering) & & 1.6 - 7.0 & 1.3 \\ \cline{2-9}
&  \multirow{2}{*}{\emph{forward-up}}& \multirow{2}{*}{$>+25$\,\AA{}} & \multirow{2}{*}{$>0$} &\multirow{2}{*}{\textbf{VII}}&\textbf{indirect emission} via elastic collision with charge transfer & \multirow{2}{*}{h, green} &1.5 - 7.1 & 0.9 \\
&&&&& and subsequent \textbf{knock up} (inelastic scattering) & & 2.5 - 4.6 &  0.04 \\ \cline{2-9}
&  \multirow{3}{*}{\emph{forward-down}}& \multirow{3}{*}{$>+25$\,\AA{}} & \multirow{3}{*}{$<0$} & \textbf{I}&\textbf{direct emission} of the weakly bound electron &f, blue & 0.7 - 5.3 & 22.0  \\ \cline{5-9}
&&&& \textbf{VIII}& \textbf{knock-down} induced charge transfer (following \textbf{VII}) &h, ---  & 4.2 - 18.5 & 0.5 \\ \cline{5-9}
&&&& \textbf{XI}& charge transfer via \textbf{non-adiabatic transition} (following \textbf{VII}) &h, green  & 9.9 - 19.3& 0.5 \\
\hline\hline
\end{tabular}
\end{table*}

Several processes (indicated by boldface roman numerals) can be identified 
and separated from each other. They 
are summarized with their associated ionization channel $S$ in Tab.\,\ref{tab:dichotomy}. 
We first consider the cases, in which the electron (here, the $x$ electron) interacting with the 
electric field is the one being eventually emitted 
({Fig.\,\ref{fig:pop}c--f}).
The strongest signal (panel f) corresponds to the emission of the weakly
bound electron, initially located near $x=+5$\,\AA{}~(subsystem B), 
in \emph{forward} direction corresponding to a direct photoemission (\textbf{I})
without scattering with the remaining electron. 
In this case, the strongly bound $y$ electron remains almost unaffected in its 
position located at the left-hand side (\emph{down}), corresponding to the electronic 
ground state of the one-electron system, $n=0$ (see Fig.\,\ref{fig:energycurves}b). 

However, the removal of the $x$ electron results in an 
increase of the $y$ electron's 
binding energy corresponding to a shake-down process. 
The energy change corresponding to a sudden electron removal 
can be estimated from the potential energy curves at $R_0$,
$\Delta E_B\approx V^\mathrm{0e}(R_0)+V^\text{2e}_0(R_0)
-V^{\mathrm{1e}}_0(R_0)-V^{\mathrm{1e}}_2(R_0)$ (where $V^\text{0e}$ 
corresponds to the repulsion energy between the movable and
 the two fixed nuclei only) and accounts for 1.8\,eV.
Comparing the relative populations at $t=5$\,fs, only a very weak 
shake-up into the one-electron states $n=1$ and 3 is noticed 
with populations of $P^\text{fwd}_1/P^\text{fwd}_0=0.0045$ 
and $P^\text{fwd}_3/P^\text{fwd}_0=0.0004$ relative to the groundstate. This
is in line with the almost constant average momentum 
seen in Fig.\,\ref{fig:py}d for the full (fermionic) system, 
indicating no coherent dynamics induced. 

In contrast, emitting the weaker bound $x$ electron in the opposite (\emph{backward}) direction,  
Fig.\,\ref{fig:pop}e, such that inelastic scattering with the remaining 
strongly-bound $y$ electron occurs, entails a significant relative population of the 
first and third excited one-electron states of 
$P^\text{bwd}_1/P^\text{bwd}_0=0.21$ and $P^\text{bwd}_3/P^\text{bwd}_0=0.02$, respectively.
Note, that these excited states are still localized on the left-hand 
side of the molecule (\emph{down} channel). 
Since such an excitation does not occur in the \emph{forward} direction, 
it must be a result of dynamical correlation between the accelerated 
$x$ electron and the "inactive" $y$ electron.
This interaction corresponds therefore to a 
pure knock-up process (\textbf{II}) \cite{Sukiasyan12PRA}.
As a consequence, within 2 and 5\,fs, a $y$ electron wave packet 
can be seen, oscillating within the left potential well, which 
is reflected in the damped oscillation pattern of the average momentum 
shown in Fig.\,\ref{fig:py}c.

A similar situation is encountered, when the stronger bound electron 
(Fig.\,\ref{fig:pop}c and d) interacts with the electric field and is 
ultimately emitted.
If the $x$ electron emission occurs in the \emph{backward} direction, 
Fig.\,\ref{fig:pop}c, again, no intramolecular scattering occurs (\textbf{III}) and
the weaker bound $y$ electron remains (initially) in its 
place, corresponding predominantly to the second excited state, $n=2$, 
which is localized on the 
molecule's right-hand side (\emph{up} channel).
Again, a shake-down stabilization of the binding energy of approximately 1.8\,eV is 
expected and only a very weak shake-up to state $n=4$ is noted ($P^\text{bwd}_4/P^\text{bwd}_2=0.0048$).

In the \emph{forward} direction, i.e.~with immediate electron-electron interaction (\textbf{IV}), 
Fig.\,\ref{fig:pop}d, the second excited 
state is dominant, too but also shows a considerable knock-up process ($P^\text{fwd}_4/P^\text{fwd}_2=0.10$).
Additionally, an increase in the population of the one-electron 
ground state, $n=0$, can be noted.
Therefore, inelastic intramolecular scattering with the weaker bound $y$
electron must have taken place resulting in a
knock-down process (\textbf{V}) of the $y$ electron
during the $x$ electron emission from the (stronger bound) lower energy levels.
Comparing their respective peaks, the relative population 
achieved through the knock-down process 
is $P^\text{fwd}_0/P^\text{fwd}_2=0.45$.
Since in the energetically lower $n=0$ state, 
the $y$ electron is located on the left-hand side, 
this correlation-driven process coincides with an intramolecular charge transfer (green arrows).

Finally, the lowest panels, Fig.\,\ref{fig:pop}g and h, correspond purely to 
correlation-driven processes, in which the energy provided by the electric 
field is absorbed by electron $y$, but results eventually in the emission 
of electron $x$. 
Therefore, a nearly elastic collision between the two electrons must 
have occurred, in which the absorbing electron ($y$) transfers most of its 
acquired kinetic energy to the electron originally unaffected by the field ($x$). 
This "indirect photoemission" process (\textbf{VI},\textbf{VII}), is 
very similar to the two-step-one (TS1) process in double ionization, 
where the electron emitted first pushes another electron 
out of an atom in a second step after photoabsorption \cite{Hino93PRA}.
But in our case, the initially accelerated electron is not released in the end
but rather takes the place of the 
subsequently emitted electron -- similar to the elastic collision between billiard balls.
Consequently, the resulting populations of the single-electron states of 
the remaining electron shown in Fig.\,\ref{fig:pop}g for subsystem A
resemble the one of subsystem B, seen in panel 
e, where the remaining electron is located at the strongly bound side.
The same applies for subsystem B's elastic collision process, Fig.\,\ref{fig:pop}h, 
which rather resembles the populations found in A's inelastic scattering, 
panel d including knock-up (within \textbf{VII}) and knock-down (\textbf{VIII})  features.
Again, this is traced back to the accelerated $y$ electron taking $x$'s 
place \emph{prior} to the emission of the $x$ electron.
Note that the indirect photoemission leads to a charge transfer 
(through elastic collision) immediately after photoabsorption. 
We therefore conclude that the significant early signals, 
seen in Fig.\,\ref{fig:contributions} in subsystem A's
 \emph{backward-down} channel (and with a lower amplitude also in subsystem B's 
\emph{forward-up} channel) correspond to the elastic 
collision process preceding the $x$ electron's emission.

\begin{figure}
\centering
\includegraphics[width=8.5cm, trim=7 18 0 0, clip]{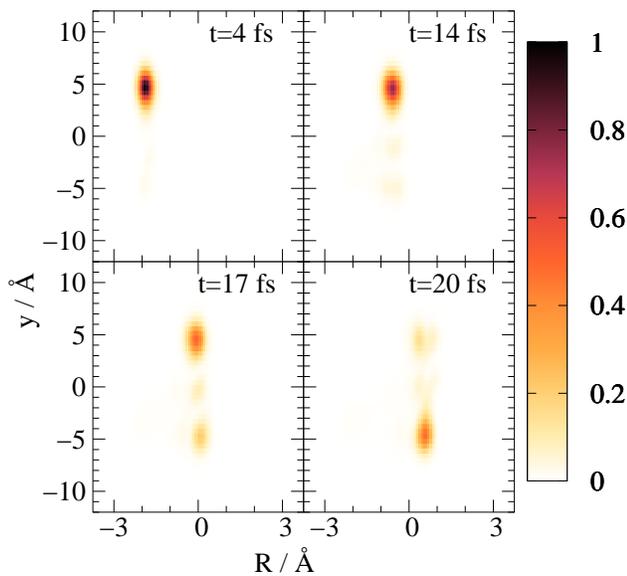}
\caption{Electron-nuclear density of subsystem A in the \emph{backward} channel, $\int \left|\psi_\mathrm{out}^{\mathrm{bwd}}(x,y,R)\right|^2\,\mathrm{d}x$. Here, the strongly bound $x$ electron is emitted and the $y$ electron remains initially on weakly bound side ($y>0$), before non-adiabatic transitions around $R=0$ lead to a charge transfer with significant electron density at the molecule's left-hand side ($y<0$).}

\label{fig:Rydyn}
\end{figure}

In panels c,d, and h of Fig.\,\ref{fig:pop}, where the $n=2$ state is predominantly occupied, 
a decrease occurs in $P^{S'}_2(t)$ after approximately $t\approx15$\,fs
with a simultaneous increase of the $n=1$ state, 
$P^{S'}_1(t)$.
This time-dependent feature can be traced back to the non-adiabatic 
nuclear reorganization dynamics (\textbf{IX,X,XI}) induced by the ionization process.
Note that these transitions do not occur, if the simulation is performed with a
 frozen nuclear configuration (dotted lines).
Fig.\,\ref{fig:Rydyn} shows the correlated electron-nuclear dynamics
for subsystem A with emission in backward direction (corresponding to 
Fig.\,\ref{fig:pop}c) through the density function 
$\int \left|\psi_\text{out}^\text{bwd}(x,y,R,t)\right|^2\,\mathrm{d}x$.
It can be seen that during the first 14\,fs, the shape of the electronic part 
only marginally changes, while the center of the nuclear distribution moves 
from $R_0=-2.05$\,\AA{} towards larger values. This dynamics is induced by the Coulomb
attraction between the remaining electron and the mobile nucleus, but is also consistent with the 
potential energy surface of the second state, $V^{\text{1e}}_2(R)$, 
see Fig.\,\ref{fig:energycurves}a upper panel. The latter one exhibits a large gradient 
towards the molecular center ($R=0$), where a coupling region with the 
first excited state, $V^{\text{1e}}_1(R)$, is found.
Indeed, at 17\,fs, the center of the nuclear distribution passes the origin and 
the electronic distribution begins to shift towards the left-hand side 
(to negative $y$ values), which is reflected in 
a slightly negative instantaneous average electronic 
momentum $\la p_y \ra^{\mathrm{backward-up}}(t)$ in the case of direct photoionization in the 
\emph{backward-up} channel, cf.~Fig.\,\ref{fig:py}a, 
which is not present in the case of frozen nuclei (red dotted line).
Therefore, upon ionization, nuclear dynamics is initiated, driving the system 
via non-adiabatic transition from the second to the first electronically 
excited state, $\varphi^{\text{1e}}_2(y;R) \rightarrow \varphi^{\text{1e}}_1(y;R)$ corresponding
to an intramolecular charge transfer. Note, however, that this process is 
significantly slower than the charge transfer process driven by electronic correlation.

It is noteworthy that, here, despite these non-adiabatic transitions, the nuclear dynamics does
not seem to affect the various ionization processes discussed above.
We attribute this to the chosen near-equilibrium initial conditions for $\chi(R)$.
Previous studies have shown, that nuclear dynamics following initial non-equilibrium configurations 
is reflected in the photoelectron momentum distribution \cite{Falge11JCP,Falge17PCCP}.
Investigations of such effects on the correlation-driven knock-up and knock-down processes
are currently under way in our workgroup.

To summarize, all observed processes, i.e.~direct photoemission with and without 
inelastic scattering leading to knock-up and knock-down transitions, 
as well as indirect photoemission through elastic collision, and non-adiabatic 
transitions, are summarized in Tab.\,\ref{tab:dichotomy} together 
with their individual amplitudes at their respective maximum and build up times, 
i.e.~the time span the corresponding wave packet requires to achieve 
approximately stable populations within the observed time window. 
These times differ for the individual processes mostly due to the different travelling 
distances (direct emission vs.~scattered and indirect emissions) for the 
ejected electron to reach the evaluation zone and because of differences
in their kinetic energies. However, the slightly longer timescales of knock-down processes 
in particular indicate a more complex electron-electron dynamics 
within the molecular system prior to the electron release.

Regarding the relative amplitudes, we note, that the correlation-driven 
pathways through elastic and inelastic 
scattering appear to be nearly of the same order of magnitude as the 
direct photoemission. 
This can be seen in Fig.\,\ref{fig:pop}a and b, where the dynamics of the 
fully antisymmetric initial wave function 
without any restrictions on the electric-field interaction (solid lines) is qualitatively 
reproduced by the artificially restricted subsystems 
(dashed lines, corresponding to the direct sum of all individual pathways 
with restricted field interaction).
Remaining discrepancies stem most likely from the missing simultaneous interaction 
of the XUV pulse with both electrons and also from the omission of interference effects 
between emitted density from the two initial localized electronic density 
distributions, which we have dropped by regarding electrons as distinguishable 
particles.

We conclude that the final state of the molecular parent ion after ionization 
depends strongly on dynamical correlation between electrons through
elastic and inelastic intramolecular scattering events preceding the electron 
emission beyond static shake-up effects.
Furthermore, we showed that such effects also
 significantly contribute to the total ionization 
probability. 
In particular, quantifying the effect of correlation-driven 
knock-up/knock-down processes and the indirect photoemission
processes on the same 
magnitude as direct photoionization processes, underlines the 
deficiencies of the commonly 
applied single active electron approximation 
as well as the sudden approximation -- even in the context of 
single-photon ionization of molecules.

\section{\label{sec:summary}Summary}

We investigated the correlated electron-electron and electron-nuclear
dynamics in a one-dimensional molecular charge-transfer model
by solving the time-dependent Schrödinger equation numerically for two
electronic and one nuclear degree of freedom.
To this end, we considered ionization of a single electron by an
ultrashort XUV pulse and investigated the exact scattering mechanism
by carefully backtracking the electron's interaction with the residual
cation and in particular with the remaining bound electron.

We introduced different theoretical approaches to investigate the correlated 
electron dynamics and nuclear motion during and after the pulse interaction. 
First, by removing all
unionized components from the total wave function and projection 
on the one-electron system to quantify shake and knock processes. 
Second, we dissected these processes and  identified 
different contributions of ionization pathways
by replacing the fermionic wave functions by that of two "distinguishable"
electrons and by restricting the interaction of the XUV pulse to a
specific electron within the molecule.
Using this approach, time-dependent signatures in the evolution of
the molecular parent ion were identified and traced back to various
intramolecular scattering events on different time scales.

Thereby, we went beyond commonly employed approximations such as 
the sudden approximation, frozen nuclear degrees of freedom, or the single active electron
approximation. Thus, significant contributions from electron-electron and 
non-adiabatic interactions within the single-photon ionization process were revealed 
on the atto- and femtosecond timescale.
In particular, relevant pathways to the overall signal were isolated,
in which inelastic scattering  resulted in knock-up and knock-down phenomena
beyond the typically regarded (sudden) shake effects.
Additional pathways of significant contribution involving elastic scattering were found, where the electron originally accelerated by the
electric field transfers its momentum to a different electron within the
molecule and takes its place instead (indirect photoemission). 
Our analysis revealed differences in the temporal signatures of all 
identified processes and allowed to estimate their relevance within the 
overall photoemission process. It was shown that electron-correlation driven processes
occur on the same order of magnitude as the direct photoemission.
While for a two-electron system the amplitude of the elastic
collision process may be overestimated due to the reduced dimensionality
of the model system, we expect this process to become
even more relevant in larger, multi-electron systems.
Furthermore, it was shown that different ionization pathways leave the
parent molecular ion in different electronic states. As a consequence,
correlated electron-nuclear reorganization dynamics is induced. 

We believe that the observations made here for a model system are
representative for molecular systems and consequently that both, 
elastic and inelastic scattering among electrons, contribute significantly to the ionization
processes and the postionization dynamics through various pathways beyond the single active electron
picture.

\begin{acknowledgments}
F.\,G.\,F., S.\,G., and U.\,P. highly acknowledge support from 
the German Science Foundation DFG, IRTG 2101. 
A.\,S.~and S.\,G.~also acknowledge the ERC Consolidator Grant QUEM-CHEM.
K.\,M.\,Z.~and S.\,G.~are part of the Max Planck School of Photonics 
supported by BMBF, Max Planck Society, and Fraunhofer Society.
\end{acknowledgments}

\end{document}